\documentclass[12pt,a4paper]{iopart}

\usepackage[english]{babel}

\usepackage{graphicx}

\usepackage{iopams}

\usepackage{bm}

\usepackage{hyperref}

\usepackage{cite}

\usepackage{multirow}

\newcommand{\ii}{\rmi}

\newcommand{\dd}{\rmd}
\newcommand{\stirling}[2]{\ensuremath{\mathcal{S}(#1,#2)}}
\newcommand{\tailor}[2]{\ensuremath{{#1}^{[#2]}}}
\newcommand{\Tailor}[2]{\ensuremath{{#1}^{(#2)}}}
\renewcommand{\Re}{{\rm Re}\,}
\renewcommand{\Im}{{\rm Im}\,}

\begin{document}

\title{Harmonic inversion analysis of exceptional points in resonance
  spectra}
\author{Jacob Fuchs, J\"org Main, Holger Cartarius, G\"unter Wunner}
\address{1. Institut f\"ur Theoretische Physik,
  Universit\"at Stuttgart, 70550 Stuttgart, Germany}

\begin{abstract}
The spectra of, e.g.\ open quantum systems are typically given as the
superposition of resonances with a Lorentzian line shape, where each
resonance is related to a simple pole in the complex energy domain.
However, at exceptional points two or more resonances are degenerate
and the resulting non-Lorentzian line shapes are related to higher
order poles in the complex energy domain.  In the Fourier-transform
time domain an $n$-th order exceptional point is characterised by a
non-exponentially decaying time signal given as the product of an
exponential function and a polynomial of degree $n-1$.  The complex
positions and amplitudes of the non-degenerate resonances can be
determined with high accuracy by application of the nonlinear harmonic
inversion method to the real-valued resonance spectra.  We extend the
harmonic inversion method to include the analysis of exceptional
points.  The technique yields, in the energy domain, the amplitudes of
the higher order poles contributing to the spectra, and, in the time
domain, the coefficients of the polynomial characterising the
non-exponential decay of the time signal.  The extended harmonic
inversion method is demonstrated on two examples, viz.\ the analysis
of exceptional points in resonance spectra of the hydrogen atom in
crossed magnetic and electric fields, and an exceptional point
occurring in the dynamics of a single particle in a time-dependent
harmonic trap.
\end{abstract}


\section{Introduction}
\label{sec:introduction}
Nonlinear methods such as filter diagonalization \cite{Wal95,Man97a}
and harmonic inversion \cite{Man97b,Bel00a,Bel00b} have been
established which allow for the high-resolution analysis of signals,
i.e., the frequencies and amplitudes of a time signal with finite
length can be determined with high accuracy without the restriction by
the uncertainty principle of the Fourier transform.
Those methods can be applied to obtain the eigenspectra of large
matrices \cite{Bel99}, semiclassical periodic orbit quantisation
\cite{Mai97b,Mai98,Mai99a,Mai99d,Mai00}, and the analysis of quantum
spectra \cite{Mai97a,Mai99c}.
It is also possible to analyse spectra given as the superposition of
resonances with Lorentzian line shape (e.g.\ the level density or
scattering cross section of open quantum systems or microcavities),
and to determine the exact positions and corresponding amplitudes of
the resonance poles in the complex energy plane \cite{Wie08,Scho09}.

So far, the harmonic inversion method has typically been applied to
spectra with non-degenerate complex resonances, and the formulae and
derivations presented, e.g.\ in \cite{Man97b,Bel00a,Mai00,Wie08} are
restricted to that case.
However, two or more resonances can degenerate at an exceptional point
(EP) \cite{Kato66} in the complex energy plane.
The occurrence of EPs in open quantum systems has attracted much
attention since at these points not only the resonance energies but
also the eigenstates coalesce \cite{Hei90,Hei08,Hei12}.
Mathematically, the EPs are branching singularities of non-Hermitian
operators \cite{Moi11} and show interesting properties such as the
permutation of states when the EP is encircled along a closed path in
the parameter space of the system.
Theoretical investigations have revealed the existence of EPs, e.g.\
in spectra of the hydrogen atom in crossed magnetic and electric
fields \cite{Car07,Car09} and in Bose-Einstein condensates with
long-range interactions \cite{Car08,Gut13}.
Experimentally, EPs have been observed in spectra of microwave
resonators \cite{Phi00,Dem01,Dem03,Dem04,Die07,Die11} and in
electronic circuits \cite{Ste04}.
Recently, an experiment has been proposed to observe a third order EP
in the dynamics of a single particle in a time-dependent harmonic trap
\cite{Uzd13}.

In most of the systems mentioned above signatures of an EP have been
verified using properties of the nearly degenerate states in the close
vicinity of the exact degeneracy, viz.\ the permutation behaviour of
states when the EP is encircled in the parameter space.
However, in some systems, e.g.\ that discussed by Uzdin \etal
\cite{Uzd13}, appropriate parameters for the encircling of an EP do
not exist or at least are not experimentally accessible.
Therefore, alternative methods for the verification of EPs are needed.
Dietz \etal \cite{Die11} have verified the existence of the EP in a
microwave billiard with time-reversal invariance violation both via
its encircling and, as an alternative, via directly observing the
coalescence of two eigenvectors at the EP.
In this paper, we study the features of EPs close to or even exactly
at the degeneracy.
In resonance spectra a resonance related to an EP is characterised by
a non-Lorentzian line shape.
In the time domain the survival probability of that resonance shows a
non-exponential decay \cite{Hei10,Car11}.
We extend the harmonic inversion method to handle the degeneracies
occurring at EPs.
The extended method provides the amplitudes of the first and higher
order poles in the complex energy or frequency plane and thus the
exact profile of both Lorentzian and non-Lorentzian resonance line
shapes.
In the time domain the harmonic inversion method yields the decay
signal of a resonance given by an exponential function multiplied by
a polynomial in time.
The observation of a resonance with non-Lorentzian line shape or
non-exponentially decaying time signal can be taken as clear evidence
for the existence of an exceptional point in the system.

The paper is organised as follows.
In section~\ref{sec:hi} the harmonic inversion method is introduced
and extended to the case of degenerate resonances.
The method is applied to the analysis of exceptional points by way of
two examples in section~\ref{sec:EP_analysis}.
Firstly, in section~\ref{sec:Hatom} the non-Lorentzian shape of a
resonance related to an EP is revealed in the photoabsorption spectrum
of the hydrogen atom in crossed magnetic and electric fields, and,
secondly, in section~\ref{sec:harm_trap} a third order EP is observed
in the dynamics of a single particle in a time-dependent harmonic trap.
Conclusions are drawn in section~\ref{sec:conclusion}.

\section{Harmonic inversion method}
\label{sec:hi}
For the convenience of the reader we first briefly recapitulate the
harmonic inversion method for spectra with non-degenerate resonances
and then extend the method by including degeneracies.

\subsection{Harmonic inversion of spectra with non-degenerate
  resonances}
\label{sec:hi_non-degenerate}
A spectrum of non-degenerate resonances is given by the imaginary
part of the function
\begin{equation}
  G(w) = \sum_k \frac{d_k}{w-w_k} \; ,
\label{eq:G-NE}
\end{equation}
where the complex parameters $w_k$ with $\Im w_k<0$ determine the
positions and widths of the resonances and the complex $d_k$ describe
the amplitudes and details of the Lorentzian line shapes.
The aim of the harmonic inversion method is to extract the parameters
$w_k$ and $d_k$ from the spectrum given on an equidistant grid of
points along the real $w$-axis.

The first step is to apply the Fourier transform to $G(w)$ to obtain
the signal
\begin{equation}
  C(s) = \frac{1}{2\pi} \int_{w_-}^{w_+} G(w) \exp\left(-\ii w s\right) \dd w
       = -\ii \sum_{k=1}^K d_k \exp \left(-\ii w_k s \right) \; .
\label{eq:C-NE}
\end{equation}
The parameters $w_\pm$ in \eref{eq:C-NE} are introduced to analyse the
spectrum in the window $[w_-,w_+]$, i.e.\ the real parts of the $w_k$
on the r.h.s.\ of \eref{eq:C-NE} are in the range $w_- < \Re w_k < w_+$.
Note that $K$ is the number of poles in that window and can be small
compared to the number of poles in the full spectrum $G(w)$.
The set of resonances in a larger region is then obtained by analysing
several consecutive windows \cite{Bel00a}.
The harmonic inversion of band-limited signals $C(s)$ with $K$ not
larger than about 50 to 200 is numerically more efficient and stable
compared to the analysis of a signal with a huge number of frequencies.
When the spectrum is given on equidistant grid points
$w=w_-+\sigma\Delta w$ for $\sigma=0,1,\dots,N-1$ the Fourier integral
in \eref{eq:C-NE} is approximated by a discrete sum.
Using the time step $\tau=2\pi/(w_+-w_-)$ the signal points
\begin{equation}
  c_n \equiv C(n\tau) = \frac{\exp \left( - \ii n\tau  w_- \right)}{2\pi}
  \sum_{\sigma=0}^{N-1} G(w_- + \sigma  \Delta w) \exp \left(
    - 2\pi\ii \frac{n\sigma}{N} \right)
\label{eq:Cntau}
\end{equation}
are easily obtained for $n=0,1,\dots,N-1$ with a fast Fourier
transform (FFT) algorithm.
Considering the factor $\exp(-\ii n\tau w_-)$ in \eref{eq:Cntau} as a
shift of resonances, $w_k \to w_k - w_-$, the comparison with the
signal $C(s=n\tau)$ in \eref{eq:C-NE} yields
\begin{equation}
  c_n \equiv C(n\tau) = \sum_{k=1}^K (-\ii d_k) \{\exp[-\ii (w_k-w_-) \tau]\}^n
  = \sum_{k=1}^K {\hat d}_k z_k^n
\label{eq:cn}
\end{equation}
for $n=0,1,\dots,N-1$ and with the abbreviations ${\hat d}_k = -\ii d_k$
and $z_k = \exp[-\ii (w_k-w_-) \tau]$.

The second step is to determine the $2K$ parameters ${\hat d}_k$ and
$z_k$ for given signal points $c_n$.
Note that at least $N=2K$ signal points are required to solve the
nonlinear set of equations \eref{eq:cn}.
As the number of frequencies in the band-limited signal is relatively
small ($K \sim 50 - 200$) several methods, which otherwise would be
numerically unstable, can be applied, e.g.\ linear predictor, Pad\'e
approximant, or direct signal diagonalization \cite{Bel00a}.
Here, we resort to the Pad\'e approximant.
Let us assume for the moment that the signal points $c_n$ are known up
to infinity, $n=0,1,\dots\infty$.
Interpreting the $c_n$'s as the coefficients of a Maclaurin series in
the variable $z^{-1}$, we can define the function
$g(z)=\sum_{n=0}^\infty c_n z^{-n}$.
With \eref{eq:cn} and the sum rule for geometric series we obtain
\begin{equation}
    g(z) \equiv \sum_{n=0}^\infty c_n z^{-n}
  = \sum_{k=1}^K \hat d_k \sum_{n=0}^\infty (z_k/z)^n
  = \sum_{k=1}^K \frac{z \hat d_k}{z-z_k} \equiv \frac{P_{K}(z)}{Q_K(z)} \; .
\label{g_Pade:eq}
\end{equation}
The right-hand side of \eref{g_Pade:eq} is a rational function with
polynomials of degree $K$ in the numerator and denominator.
Evidently, the parameters $z_k$ are the poles of $g(z)$, i.e., the
zeros of the polynomial $Q_K(z)$.
Of course, the assumption that the coefficients $c_n$ are known up to
infinity is not fulfilled and, therefore, the sum over all $c_nz^{-n}$
in \eref{g_Pade:eq} cannot be evaluated in practice.
However, the convergence of the sum can be accelerated by use of
the Pad\'e approximant.
Indeed, knowledge of $2K$ signal points $c_0,\dots,c_{2K-1}$ is 
sufficient for the calculation of the coefficients of the two polynomials
\begin{equation}
 P_{K}(z) = \sum_{k=1}^{K} b_k z^k  \mbox{~~and~~}
 Q_K(z) = \sum_{k=1}^K a_k z^k - 1 \; .
\label{P_Q_polynomials:eq}
\end{equation}
The coefficients $a_k$ with $k=1,\dots,K$ are obtained as solutions of
the linear set of equations
\begin{equation}
 c_n = \sum_{k=1}^K a_k c_{n+k} \; ,  \quad n = 0, \dots, K-1 \; .
\end{equation}
Once the $a_k$ are known, the coefficients $b_k$ are given by the
explicit formula
\begin{equation}
 b_k = \sum_{m=0}^{K-k} a_{k+m} c_{m} \; , \quad k = 1, \dots , K \; .
\label{eq:bk}
\end{equation}
The parameters $z_k=\exp[-\ii (w_k-w_-)\tau]$ and thus the frequencies
\begin{equation}
 w_k = w_- + \frac{\ii}{\tau}\ln{z_k}
\label{eq:wk}
\end{equation}
(choosing the branch of the logarithm with $0\le\Im\ln z<2\pi$)
are obtained by searching for the zeros of the polynomial $Q_K(z)$ in
\eref{P_Q_polynomials:eq}.
This is the only nonlinear step of the algorithm.
The roots of polynomials can be found, in principle, by application
of Laguerre's method \cite{Pre92}.
However, it turns out that an alternative method, i.e.\ the 
diagonalization of the Hessenberg matrix 
\begin{equation}
 {\bf A} =
 \left(\begin{array}{ccccc}
  -\frac{a_{K-1}}{a_K} & -\frac{a_{K-2}}{a_K} & \cdots &
  -\frac{a_{1}}{a_K} & \frac{1}{a_K} \\
  1 & 0 & \cdots & 0 & 0 \\
  0 & 1 & \cdots & 0 & 0 \\
  \vdots & & & & \vdots \\
  0 & 0 & \cdots & 1 & 0 
 \end{array} \right) \; ,
\label{Hesse:eq}
\end{equation}
with $a_k$ the coefficients of the polynomial $Q_K(z)$ in
\eref{P_Q_polynomials:eq}, is a numerically more robust technique for
finding the roots of high degree ($K \gtrsim 60$) polynomials \cite{Pre92}.

The parameters $\hat d_k$ are calculated via the residues of the last
two terms of \eref{g_Pade:eq}.
Application of the residue calculus yields
\begin{equation}
 \hat d_k = -\ii d_k = \frac{P_{K}(z_k)}{z_k Q_K'(z_k)} \; ,
\label{dk_pade:eq}
\end{equation}
with the prime indicating the derivative ${\dd}/{{\dd}z}$.
Note that $Q_K'(z_k)$ vanishes if $z_k$ is a multiple root of the
polynomial $Q_K(z)$ and thus equation \eref{dk_pade:eq} is not valid
for degenerate resonances occurring at exceptional points.
In the following we extend the harmonic inversion method to include
that case.

\subsection{Extension of the harmonic inversion method to
  degenerate resonances}
\label{sec:hi_degenerate}
Degenerate resonances cannot be described by simple poles of the
function $G(w)$ but higher order poles are required.
Therefore, the ansatz \eref{eq:G-NE} is generalised to
\begin{equation}
  G(w) = \sum_k \sum_{\alpha=1}^{r_k} \frac{d_{k,\alpha}}{(w-w_k)^{\alpha}} \; ,
\label{eq:spektrum}
\end{equation}
where $r_k$ is the order of degeneracy of resonance $k$.
Note that resonances with $r_k>1$ in spectra given as $\Im G(w)$ have
a non-Lorentzian line shape.
The aim is now to extract the complex resonance positions $w_k$ and
amplitudes $d_{k,\alpha}$ from a given spectrum.

The first step is again the computation of a band-limited time signal
by a windowed Fourier transform of $G(w)$.
Writing $G(w)$ in \eref{eq:spektrum} as
\begin{equation*}
  G(w) = \sum_k \sum_{\alpha=1}^{r_k} \; d_{k,\alpha}
  \frac{(-1)^{\alpha-1}}{(\alpha-1)!}
  \left({\textstyle\frac{\dd}{\dd w}}\right)^{\alpha-1} (w-w_k)^{-1}
\end{equation*}
we obtain
\begin{eqnarray}
  C(s) &=& \frac{1}{2\pi} \int_{w_-}^{w_+} G(w) \exp\left(-\ii
    ws\right) \dd w \nonumber \\
  &=& \sum_k \sum_{\alpha=1}^{r_k} \; d_{k,\alpha} \; \frac{(-1)^{\alpha-1}}{(\alpha-1)!} \; (\ii s)^{\alpha-1} (-\ii) \exp\left(-\ii w_k s\right) \; ,
\label{eq:fouriertransformierte}
\end{eqnarray}
where the sum over $k$ is now restricted to those resonances with
$\Re w_k \in [w_-,w_+]$.
At discrete grid points $s = n\tau$ (with $n=0,1,\dots,N-1$) the
signal reads
\begin{equation}
  c_n \equiv C(n\tau) = \sum_k \sum_{\alpha=1}^{r_k} \; (-\ii
  d_{k,\alpha}) \frac{(-\ii n\tau)^{\alpha-1}}{(\alpha-1)!}
  \exp[-\ii (w_k-w_-) n\tau] \; .
\end{equation}
With the abbreviations
\begin{equation}
  \hat d_{k,\alpha} = -\ii d_{k,\alpha} \frac{(-\ii\tau)^{\alpha-1}}{(\alpha-1)!}
\mbox{~~and~~}
  z_k = \exp[-\ii (w_k-w_-) \tau]
\end{equation}
we finally obtain the nonlinear set of equations
\begin{equation}
  c_n = \sum_k \sum_{\alpha=1}^{r_k} \; \hat d_{k,\alpha} n^{\alpha-1} z_k^n
  \quad , \quad n=0,1,\dots,2K-1 \; ,
\label{eq:cn_degenerate}
\end{equation}
where $K=\sum_k r_k$ is the total number of resonances of the
band-limited signal counting all multiplicities.

The second step, i.e.\ the solution of the nonlinear set of equations
\eref{eq:cn_degenerate}, also starts analogously to the non-degenerate
case.
Interpreting the $c_n$'s in \eref{eq:cn_degenerate} as the
coefficients of a Maclaurin series in $z^{-1}$ given for
$n=0,1,\dots,\infty$ we now obtain
\begin{equation}
    g(z) \equiv \sum_{n=0}^\infty c_n z^{-n}
  = \sum_k \sum_{\alpha=1}^{r_k} \; \hat d_{k,\alpha}
        \sum_{n=0}^\infty n^{\alpha-1} (z_k/z)^n
  \equiv \frac{P_{K}(z)}{Q_K(z)} \; .
\label{eq:g_degenerate}
\end{equation}
The rational function with the polynomials $P_{K}(z)$ and $Q_K(z)$ on
the r.h.s.\ of \eref{eq:g_degenerate} is obtained in the same way as
in \eref{g_Pade:eq} using equations \eref{P_Q_polynomials:eq}-\eref{eq:bk}.
The parameters $z_k$ are the roots of the polynomial $Q_K(z)$,
however, the difference to the non-degenerate case is that now each
root $z_k$ is $r_k$-fold degenerate.
For $\alpha > 1$ the sum on the l.h.s.\ of \eref{eq:g_degenerate} is
no longer a simple geometric series.
In the following we use the relation (see appendix)
\begin{equation}
  \sum_{n=0}^{\infty} n^{\alpha-1} \, x^n = \sum_{n=0}^{\alpha-1} n! \,
  \stirling{\alpha-1}{n} \, \frac{x^n}{(1-x)^{n+1}}
  = \frac{x^\alpha f_{\alpha}(x)}{(x-1)^{\alpha}}
\label{eq:sum_relation}
\end{equation}
with
\begin{equation}
  \stirling{n}{k} = \frac{1}{k!} \sum_{\mu=0}^{k} (-1)^{k-\mu}
  {k \choose \mu} \mu^n
\end{equation}
the Stirling numbers of the second kind \cite{Abr64}, and the
functions
\begin{eqnarray}
  f_{\alpha}(x) 
  &=& (-1)^{\alpha} \sum_{\nu=0}^{\alpha-1} \nu! \,
  \stirling{\alpha-1}{\nu} \,
  \frac{1}{x}\left(\frac{1-x}{x}\right)^{\alpha-1-\nu} \; .
\label{eq:f_alpha}
\end{eqnarray}
A few Stirling numbers $\stirling{n}{k}$ for $n,k \leq 5$ are given in
table \ref{tab1}.
\begin{table}
\centering
\caption{Stirling numbers of the second kind $\stirling{n}{k}$ for
  $n,k \leq 5$.}
\label{tab1}
\begin{tabular}{rrrrrrr}
& \multicolumn{6}{c}{$k$} \\
\multicolumn{1}{c}{$\stirling{n}{k}$} & 0 & 1 & 2 & 3 & 4 & 5 \\
\hline
$n = 0$ & ~1 &    &    &    &    &    \\
$n = 1$ &  0 & ~1 &    &    &    &    \\
$n = 2$ &  0 &  1 &  1 &    &    &    \\
$n = 3$ &  0 &  1 &  3 &  1 &    &    \\
$n = 4$ &  0 &  1 & 17 &  6 &  1 &    \\
$n = 5$ &  0 &  1 & 15 & 25 & 10 & ~1 \\
\hline
\end{tabular}
\end{table}
Using the second term of \eref{eq:sum_relation} the function $g(z)$
can be written as
\begin{equation}
  g(z) = \sum_k \sum_{\alpha = 1}^{r_k}
  \hat{d}_{k,\alpha}  \sum_{n=0}^{\alpha-1} n! \,
  \stirling{\alpha-1}{n} \, \frac{z z_k^n}{(z-z_k)^{n+1}}
  \equiv \frac{P_{K}(z)}{Q_K(z)} \; .
\label{eq:g_degenerate2}
\end{equation}
Both sides of \eref{eq:g_degenerate2} are rational functions in $z$
with degree $K$ polynomials in the numerator and denominator.
Using a partial fraction decomposition of $P_{K}(z)/[zQ_K(z)]$ the
parameters $\hat d_{k,\alpha}$ can, in principle, be obtained as
solutions of sets of linear equations comparing the coefficients of
terms $(z-z_k)^{-n}$ with the same root $z_k$ and order $n$.
However, it is also possible to derive explicit formulae for the
computation of the $\hat d_{k,\alpha}$ \cite{Fuchs13}.
Using the last term of \eref{eq:sum_relation} equation
\eref{eq:g_degenerate2} can also be written as (we drop the subscript
$K$ on the polynomials $P(z)$ and $Q(z)$ in the following)
\begin{equation}
  g(z) = \sum_k \sum_{\alpha = 1}^{r_k} \hat d_{k,\alpha}
  \frac{f_{k,\alpha}(z)}{(1-z/z_k)^\alpha}
  \equiv \frac{P(z)}{Q(z)}
\label{eq:g_degenerate3}
\end{equation}
with the functions $f_{k,\alpha}(z) = f_\alpha(z_k/z)$, which are
polynomials of degree $\alpha$ in $z$.
Equation \eref{eq:g_degenerate3} is now multiplied with $Q(z)$, and
both sides are expanded, for each root $z_k$, in a Taylor series of
degree $r_k-1$ around $z=z_k$.
Comparison of the coefficients of terms $(z-z_k)^{l}$ for $l=0,\dots,r_k-1$
yields
\begin{equation}
\fl
  \sum_{\alpha=r_k-l}^{r_k} \hat d_{k,\alpha} \sum_{\nu=0}^{l+\alpha-r_k}
  (r_k+\nu-l-1)!\, \stirling{\alpha}{r_k+\nu-l} \,
  z_k^{r_k+\nu-l} \, \tailor{Q}{r_k+\nu}(z_k) = \tailor{P}{l}(z_k)
\label{eq:taylor_exp}
\end{equation}
with the notation
\begin{equation}
  \tailor{f}{n}(x) = \frac{1}{n!} \, \Tailor{f}{n}(x)
  = \frac{1}{n!} \left. \left({\textstyle\frac{\dd}{\dd z}}\right)^n
 f(z)\right|_{z=x} \; .
\end{equation}
From \eref{eq:taylor_exp} we finally obtain
\begin{equation}
\fl
  \hat{d}_{k,\alpha} = \frac{\tailor{P}{r_k-\alpha}(z_k)}{(\alpha-1)!
    \, z_k^{\alpha} \, \tailor{Q}{r_k}(z_k)} -
  \sum_{\mu=\alpha+1}^{r_k} \hat{d}_{k,\mu} \, \sum_{\nu=0}^{\mu-\alpha}
  \frac{(\alpha-1+\nu)!\, \stirling{\mu}{\alpha+\nu}
  z_k^{\nu} \, \tailor{Q}{r_k+\nu}(z_k)}{(\alpha-1)! \,
  \tailor{Q}{r_k}(z_k)} \; .
\label{eq:amplituden-d-dach}
\end{equation}
For every index $k$ equation \eref{eq:amplituden-d-dach} can be
evaluated recursively starting with the highest value $\alpha=r_k$
down to $\alpha=1$.
In the following some special cases of \eref{eq:amplituden-d-dach} are
given explicitly.
For a non-degenerate resonance ($r_k = 1$) we obtain the already known
formula
\begin{equation}
  \hat{d}_{k,1} = \frac{P(z_k)}{z_k \, \tailor{Q}{1}(z_k)}
  = \frac{P(z_k)}{z_k \, \Tailor{Q}{1}(z_k)} \; .
\label{eq:d_nondeg}
\end{equation}
For a twofold degenerate resonance ($r_k = 2$) equation
\eref{eq:amplituden-d-dach} yields
\begin{eqnarray}
\fl
  \hat{d}_{k,2} &=& \frac{P(z_k)}{z_k^{2} \, \tailor{Q}{2}(z_k)} =
  \frac{2 \, P(z_k)}{z_k^{2} \, \Tailor{Q}{2}(z_k)} \; , \nonumber \\
\fl
  \hat{d}_{k,1}
  &=& \frac{\tailor{P}{1}(z_k)}{z_k \, \tailor{Q}{2}(z_k)} -
  \hat{d}_{k,2} \left( 1 + \frac{z_k \,
      \tailor{Q}{3}(z_k)}{\tailor{Q}{2}(z_k)} \right)
  = \frac{2 \Tailor{P}{1}(z_k)}{z_k \Tailor{Q}{2}(z_k)} -
  \hat{d}_{k,2} \left( 1 + \frac{z_k \Tailor{Q}{3}(z_k)}{3
      \Tailor{Q}{2}(z_k)} \right) \; .
\label{eq:d_twofold}
\end{eqnarray}
Threefold degenerate resonances will probably occur more rarely.
For $r_k=3$ the amplitudes read:
\begin{eqnarray}
\fl
  \hat{d}_{k,3} &=& \frac{P(z_k)}{2 z_k^{3} \tailor{Q}{3}(z_k)} =
  \frac{3 P(z_k)}{z_k^{3} \Tailor{Q}{3}(z_k)} \; , \nonumber \\
\fl
  \hat{d}_{k,2} 
  &=& \frac{\tailor{P}{1}(z_k)}{z_k^{2} \tailor{Q}{3}(z_k)} -
  \hat{d}_{k,3} \left( 3 + \frac{2 z_k
      \tailor{Q}{4}(z_k)}{\tailor{Q}{3}(z_k)} \right)
  = \frac{6 \Tailor{P}{1}(z_k)}{z_k^{2} \Tailor{Q}{3}(z_k)} -
  \hat{d}_{k,3} \left( 3 + \frac{z_k \Tailor{Q}{4}(z_k)}{2
      \Tailor{Q}{3}(z_k)} \right) \; , \nonumber \\
\fl
  \hat{d}_{k,1}
  &=& \frac{\tailor{P}{2}(z_k)}{z_k  \tailor{Q}{3}(z_k)}
  - \hat{d}_{k,2}  \left( 1 + \frac{z_k 
      \tailor{Q}{4}(z_k)}{\tailor{Q}{3}(z_k)} \right)
  - \hat{d}_{k,3}  \left( 1 + \frac{3  z_k 
      \tailor{Q}{4}(z_k)}{\tailor{Q}{3}(z_k)} + \frac{2  z_k^{2} 
      \tailor{Q}{5}(z_k)}{\tailor{Q}{3}(z_k)} \right) \nonumber \\
\fl
  &=& \frac{3  \Tailor{P}{2}(z_k)}{z_k  \Tailor{Q}{3}(z_k)}
  - \hat{d}_{k,2}  \left( 1 + \frac{z_k  \Tailor{Q}{4}(z_k)}{4 
      \Tailor{Q}{3}(z_k)} \right)
  - \hat{d}_{k,3}  \left( 1 + \frac{3  z_k 
      \Tailor{Q}{4}(z_k)}{4  \Tailor{Q}{3}(z_k)} + \frac{z_k^{2} 
      \Tailor{Q}{5}(z_k)}{10  \Tailor{Q}{3}(z_k)} \right) \; .
\label{eq:d_threefold}
\end{eqnarray}
The parameters $z_k$ and $r_k$ are the roots and corresponding
multiplicities of the polynomial $Q(z)$.
The values of the $w_k$ and $d_{k,\alpha}$ in \eref{eq:spektrum} are
finally obtained by using \eref{eq:wk} for the frequencies and
\begin{equation}
  d_{k,\alpha} = \ii \, \hat{d}_{k,\alpha} \,
  \frac{(\alpha-1)!}{(-\ii \tau)^{\alpha-1}}
\label{eq:d-aus-d-dach}
\end{equation}
for the amplitudes.

In practical applications of the harmonic inversion method the roots
$z_k$ of the polynomial $Q(z)$ can again be computed by e.g.\ Laguerre's
method or diagonalization of the Hessenberg matrix \eref{Hesse:eq}.
Numerically two or more roots will typically not coincide exactly,
however, the distance between neighbouring roots can become very small.
In that case, the application of formulae for the degenerate case
derived in this section is meaningful.
Close to a degeneracy related to an exceptional point the amplitudes
$d_k$ in \eref{eq:G-NE} of the non-degenerate resonances can become
very large and diverge at the exact EP \cite{Car09}.
By contrast, the amplitudes $d_{k,\alpha}$ in \eref{eq:spektrum} of
higher order poles describing the non-Lorentzian shape of a resonance
close to an EP only weakly depend on the distances between nearly
degenerate $z_k$ values, as will be shown in section
\ref{sec:EP_analysis}.
In an implementation of the extended harmonic inversion method it is
useful to introduce a parameter $\delta$ and to assume $z_k$ values
with distances less than $\delta$ to be degenerate.
The formulae for the amplitudes of a degenerate resonance require
inserting the parameter $z_k$ of the degenerate frequency.
In the numerical procedure that $z_k$ value is taken as the geometric
mean of the individual nearly degenerate $z_k$'s.

\section{Analysis of exceptional points}
\label{sec:EP_analysis}
We now present applications of the extended harmonic inversion method
for the analysis of exceptional points.
In section \ref{sec:Hatom} we analyse photoabsorption spectra of the
hydrogen atom in crossed magnetic and electric fields, and in section
\ref{sec:harm_trap} we investigate the occurrence of an EP in the
dynamics of a single particle in a time-dependent harmonic trap.

\subsection{Hydrogen atom in crossed magnetic and electric fields}
\label{sec:Hatom}
Atoms in crossed static magnetic and electric fields are open quantum
systems, which have been investigated in detail both experimentally
\cite{Wie89,Rai94,Fre02,Sta05} and theoretically
\cite{Mai92,Jaf00,Car09b,Car10}.
For the hydrogen atom in crossed fields the Hamiltonian [in atomic
units with $\gamma=B/(2.35\times 10^5\,{\rm T})$, $f=F/(5.14\times
10^{11}\, {\rm V/m})$, and $L_z$ the $z$-component of the angular
momentum] reads
\begin{equation}
  H = \frac{1}{2} \, \boldsymbol{p}^2 - \frac{1}{r} + \frac{1}{2} \, \gamma
  L_z + \frac{1}{8} \, \gamma^2 (x^2+y^2) + f x \; .
\end{equation}
As the crossed fields hydrogen atom is an open system the
photoabsorption spectrum is a superposition of resonances with nonzero
line widths.
The photoabsorption cross section for excitations of an initial state
$\psi_0$ at energy $E_0$ to a final state at energy $E$ can be written
as \cite{Res75}
\begin{equation}
  \sigma(E) = 4\pi\alpha (E-E_0) \, \Im
  \left(\sum_j \frac{\langle\psi_0|D|\psi_j\rangle^2}{E_j-E} \right) \; ,
\label{eq:sigma_E}
\end{equation}
with $\alpha$ the fine-structure constant and $D$ the dipole operator.
The $E_j$ are complex energy eigenvalues and the $\psi_j$ the
corresponding complex eigenstates, which can be computed with the
complex coordinate rotation method \cite{Rei82,Ho83,Mai94}.

The resonance positions in the complex energy plane depend on two
parameters, viz.\ the magnetic and electric field strengths $\gamma$
and $f$.
By varying the two field strengths it can happen that two resonances
coincide at the same complex energy.
Various methods can be applied to investigate whether or not the
degeneracy is an exceptional point.
The most established technique is to encircle the critical point in
the $(\gamma,f)$ parameter space \cite{Car07,Car09}.
The EP is a branching singularity characterised by the property that
the two states permute when the EP is encircled along a closed path.
The method, however, requires the analysis of a sufficiently high
number of measured or computed photoabsorption spectra at $(\gamma,f)$
parameters surrounding the critical point.
An alternative method is to analyse a degenerate or nearly degenerate
resonance with the extended harmonic inversion method.
Here, the fingerprint of an EP is a non-Lorentzian line shape of the
resonance caused by higher order poles ($\alpha \geq 2$) in the ansatz
\eref{eq:spektrum}.
The advantage of the method is that a single spectrum is sufficient to
reveal exceptional points.

\begin{figure}
\centering
\includegraphics[width=0.75\textwidth]{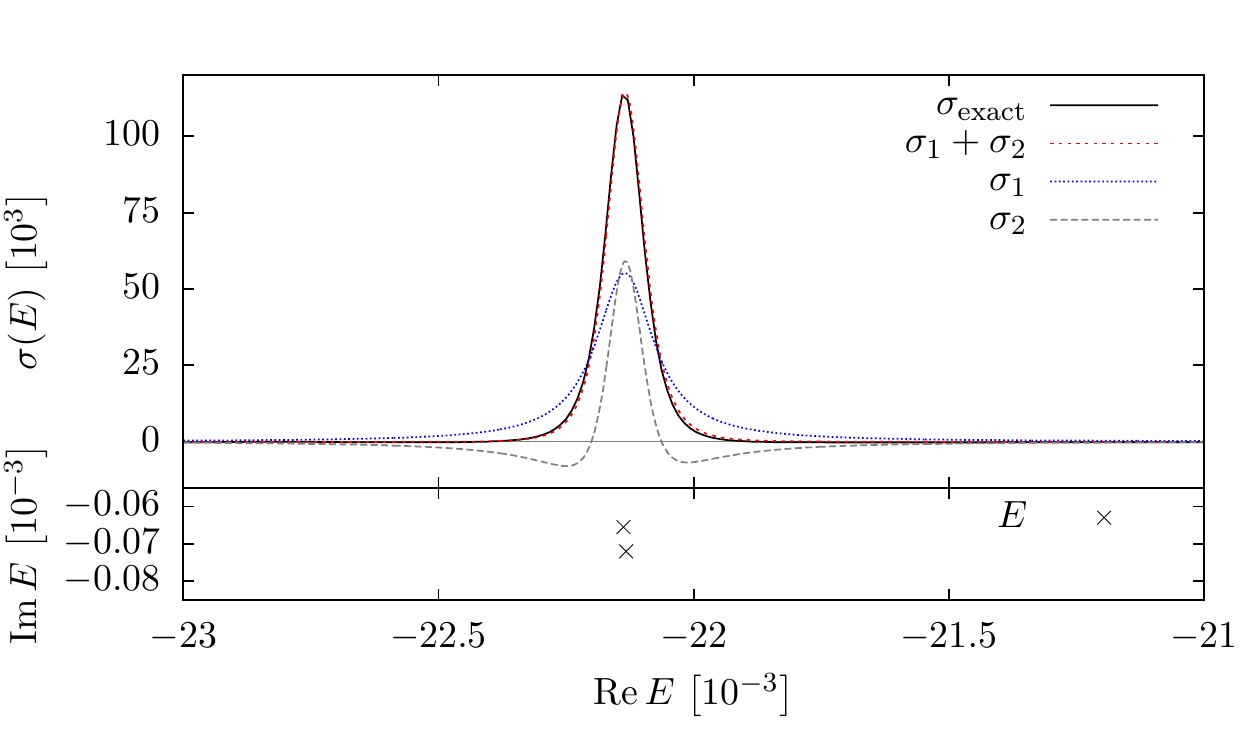}
\caption{Photoabsorption spectrum around the nearly degenerate
  resonance at $E\approx -0.0221$ of the crossed fields hydrogen
  atom at field strengths $\gamma=0.004604$ and $f=0.0002177$.
  The numerically exact cross section $\sigma_{\rm exact}$ (black
  solid line) is excellently reproduced by the sum $\sigma_1+\sigma_2$
  (red dashed line) of the contributions $\sigma_1$ related to a first
  order pole (blue dotted line) and $\sigma_2$ related to a second
  order pole (grey dashed line).  The non-vanishing contribution of
  the second order pole clearly indicates that the resonance is related
  to an exceptional point.  The positions $E_1$ and $E_2$ of the two
  nearly degenerate resonance poles are marked in the lower panel.
  All values are in atomic units.}
\label{fig1}
\end{figure}

As an example we analyse the resonance at energy $E\approx -0.0221$ in
the photoabsorption spectrum of the crossed fields hydrogen atom at
the field strengths $\gamma=0.004604$ and $f=0.0002177$ shown as solid
black line in figure~\ref{fig1}.
The photoabsorption cross section $\sigma(E)$ is computed for
transitions from the initial state $|2p0\rangle$ to final states with
light polarised parallel to the magnetic field axis as described in
\cite{Car09}.
The harmonic inversion analysis of the numerically exact spectrum
reveals two nearly degenerate complex energies $E_1$ and $E_2$ shown
in the lower panel of figure~\ref{fig1}.
They are related to the exceptional point labelled 8 in table I of
\cite{Car09}.
The values of the energies and amplitudes obtained by the harmonic
inversion method are presented in table~\ref{tab2}.
\begin{table}
\centering
\caption{Harmonic inversion analysis of the degenerate resonance at
  $E \approx -0.0221$ in figure \ref{fig1}.  The amplitudes $d_1$
  and $d_2$ have been obtained with \eref{eq:d_nondeg} for the two
  slightly different complex energies $E_1$ and $E_2$.  The amplitudes
  $d_{1,1}$ and $d_{1,2}$ have been computed with \eref{eq:d_twofold}
  for a twofold degenerate resonance at the mean energy $\bar
  E_1=(E_1+E_2)/2$.}
\label{tab2}
\begin{tabular}{lrr|lrr}
    & \multicolumn{1}{c}{$\Re E$} & \multicolumn{1}{c}{$\Im E$}
  & & \multicolumn{1}{c}{$\Re d$} & \multicolumn{1}{c}{$\Im d$} \\
\hline
  $E_1$ & -0.0221376 & -0.0000655 & $d_1$ &  10.034060 &  14.122197 \\
  $E_2$ & -0.0221322 & -0.0000720 & $d_2$ & -10.029165 & -12.224046 \\
\hline
  \multirow{2}{*}{$\bar E_1$} &
  \multirow{2}{*}{-0.0221349} &
  \multirow{2}{*}{-0.0000688} & $d_{1,1}$ &  0.004687 &  1.897540 \\
                          & & & $d_{1,2}$ & -0.000140 & -0.000006 \\
\hline
\end{tabular}
\end{table}
As the complex energies $E_1$ and $E_2$ do not exactly coincide the
two corresponding amplitudes $d_1$ and $d_2$ can be computed with
equation \eref{eq:d_nondeg} valid for non-degenerate resonances.
However, it should be noted that roughly $d_2\approx -d_1$, i.e.\ both
amplitudes cancel each other to a large amount.
The values of $d_1$ and $d_2$ strongly depend on the distance between
the two nearly degenerate resonances and diverge at the exact EP where
$E_1=E_2$.
It is therefore more appropriate to describe the resonance in
figure~\ref{fig1} using the ansatz \eref{eq:spektrum} for a
non-Lorentzian line shape.
The amplitudes $d_{1,1}$ and $d_{1,2}$ in table~\ref{tab2} have been
computed with \eref{eq:d_twofold} for a twofold degenerate resonance
at the mean energy $\bar E_1=(E_1+E_2)/2$.
The Lorentzian part of the resonance related to a first order pole in
\eref{eq:spektrum} and the non-Lorentzian part related to a second
order pole in \eref{eq:spektrum} are shown as grey dashed and blue
dotted line in figure~\ref{fig1}, respectively, and the sum of the
two contributions (red dashed line) reproduces the total
photoabsorption cross section (black solid line) very well.
The amplitudes $d_{1,1}$ and $d_{1,2}$ only weakly depend on how
closely the exact EP is approached and, in particular, do not diverge
at the EP.
This clearly demonstrates that the extended harmonic inversion method
introduced in section \ref{sec:hi_degenerate} is not restricted to the
case of exact degeneracies but is well suited even for the analysis of
nearly degenerate resonances.
Furthermore, the method allows for the detection of an EP using a
single spectrum.
The contributions of both the first and second order pole in
figure~\ref{fig1} are of similar size.
The non-vanishing contribution of the second order pole provides clear
evidence that the two resonance poles in the lower panel of
figure~\ref{fig1} are related to a second order exceptional point.
By contrast, the method of encircling the EP requires several spectra
at various parameter values to observe the permutation of states as a
fingerprint of the EP \cite{Car07,Car09}.

The spectrum in figure~\ref{fig1} has been computed numerically,
however, it is important to note that high-resolution spectroscopy of
atoms in external fields allows for the experimental observation of
exceptional points.
The resolution of the experiment must be sufficiently high to resolve
the line shapes of resonances and, in particular, to distinguish
between Lorentzian and non-Lorentzian profiles.
The harmonic inversion analysis of experimental high-resolution
spectra is an alternative to the measurement of the survival
probability $S(t) = |\langle\psi(0)|\psi(t)\rangle|^2$ of a decaying
degenerate resonance $\psi(t)$, which has been excited with a laser
whose bandwidth is large compared to the width of the resonance
\cite{Car11}.
Nevertheless, the extended harmonic inversion method can also be
applied to adjust the measured survival probability to the functional
form of the signal \eref{eq:fouriertransformierte}, and to detect,
e.g., a second order EP by the non-exponential decay of the survival
probability $S(t) = |1-at|^2\exp(2\Im E_{\rm EP} t/\hbar)$ with a
parameter $a \ne 0$.

\subsection{Single particle in a time-dependent harmonic trap}
\label{sec:harm_trap}
\begin{figure}
\centering
\includegraphics[width=0.75\textwidth]{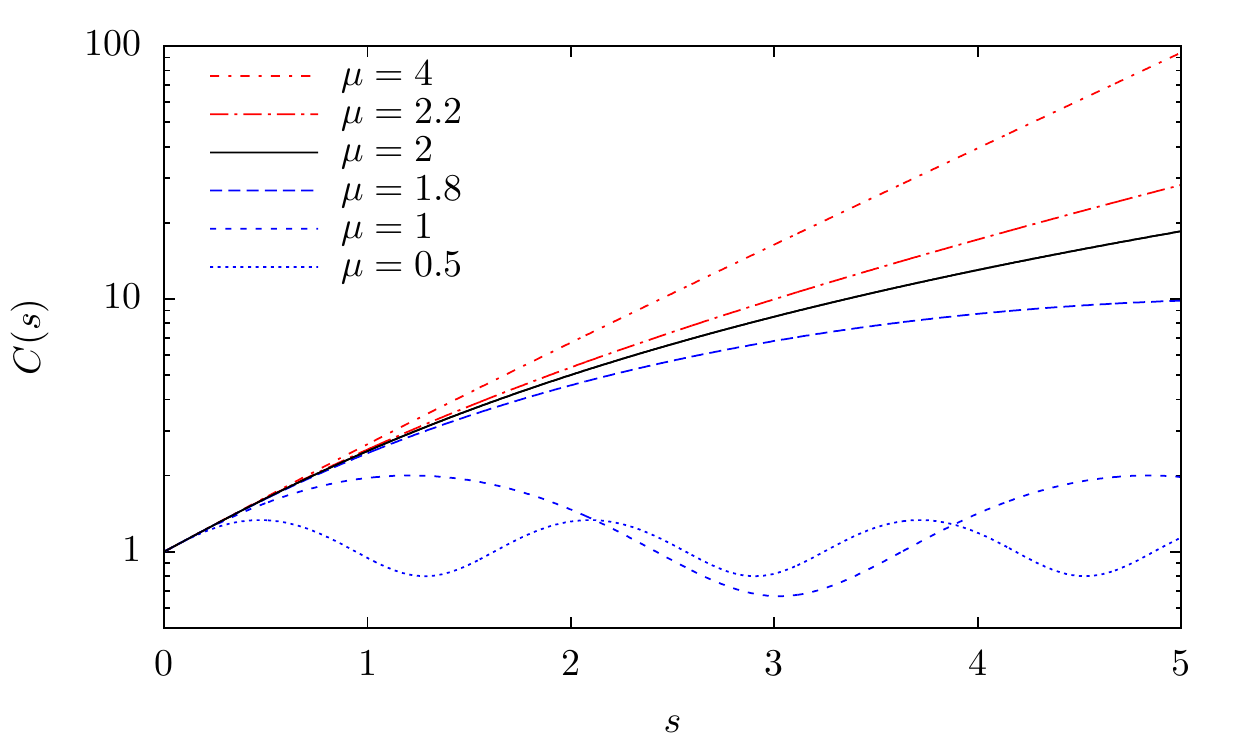}
\caption{Signal $C(s)$ obtained as normalised variance of the position
  operator $\langle x^2\rangle$ as a function of renormalised time of
  a particle in a time-dependent harmonic trap for various values of
  the adiabatic parameter $\mu$.  A transition from an oscillatory to
  a monotonically growing exponential dynamics occurs at $\mu=2$.}
\label{fig2}
\end{figure}

As a second example for the application of the extended harmonic
inversion method we study the dynamics of a single particle with mass
$m$ in a harmonic oscillator with changing frequency given by the
Hamiltonian
\begin{equation}
  H = \frac{p^2}{2m} + \frac{m}{2}\omega^2(t) x^2 \; .
\end{equation}
It has been shown by Uzdin \etal \cite{Uzd13} that when the frequency
is changed by keeping the dimensionless adiabatic parameter
\begin{equation}
  \mu = \left[\frac{1}{\omega^2(t)}\right]\frac{\dd\omega(t)}{\dd t}
\end{equation}
fixed, the time evolution features an exceptional point at $\mu=2$.
The variance of the position operator $\langle x^2\rangle$ normalised
by the width of the instantaneous potential $1/\sqrt{2m\omega(t)}$ as
function of the renormalised time $s=(1/\mu)\ln[\omega(t)/\omega(0)]$
is given by the signal
\begin{equation}
  C(s) = \frac{\mu^2 \cosh(s\sqrt{\mu^2-4}/\mu) 
  + \mu\sqrt{\mu^2-4}\sinh(s\sqrt{\mu^2-4}/\mu) - 4}{\mu^2-4}
\label{eq:C_harm_trap}
\end{equation}
and illustrated for various values of $\mu$ in figure~\ref{fig2}.
The signal $C(s)$ shows a sharp transition from an oscillatory to a
monotonic exponential dynamics at $\mu=2$, where three frequencies of
the signal are identical.
Is this degeneracy a third order EP?
That question cannot be answered with the established method of
observing the permutation of the frequencies when the EP is encircled
in the parameter space because the single control parameter $\mu$ of
the system does not allow for such an encircling.
Uzdin \etal \cite{Uzd13} propose to observe the third order derivative
of the renormalised variance which should vanish at $\mu=2$, where
$C(s)$ is a second order polynomial in $s$.
However, higher order derivatives of the experimentally measured
signal may strongly suffer from noise.

\begin{figure}
\centering
\includegraphics[width=0.75\textwidth]{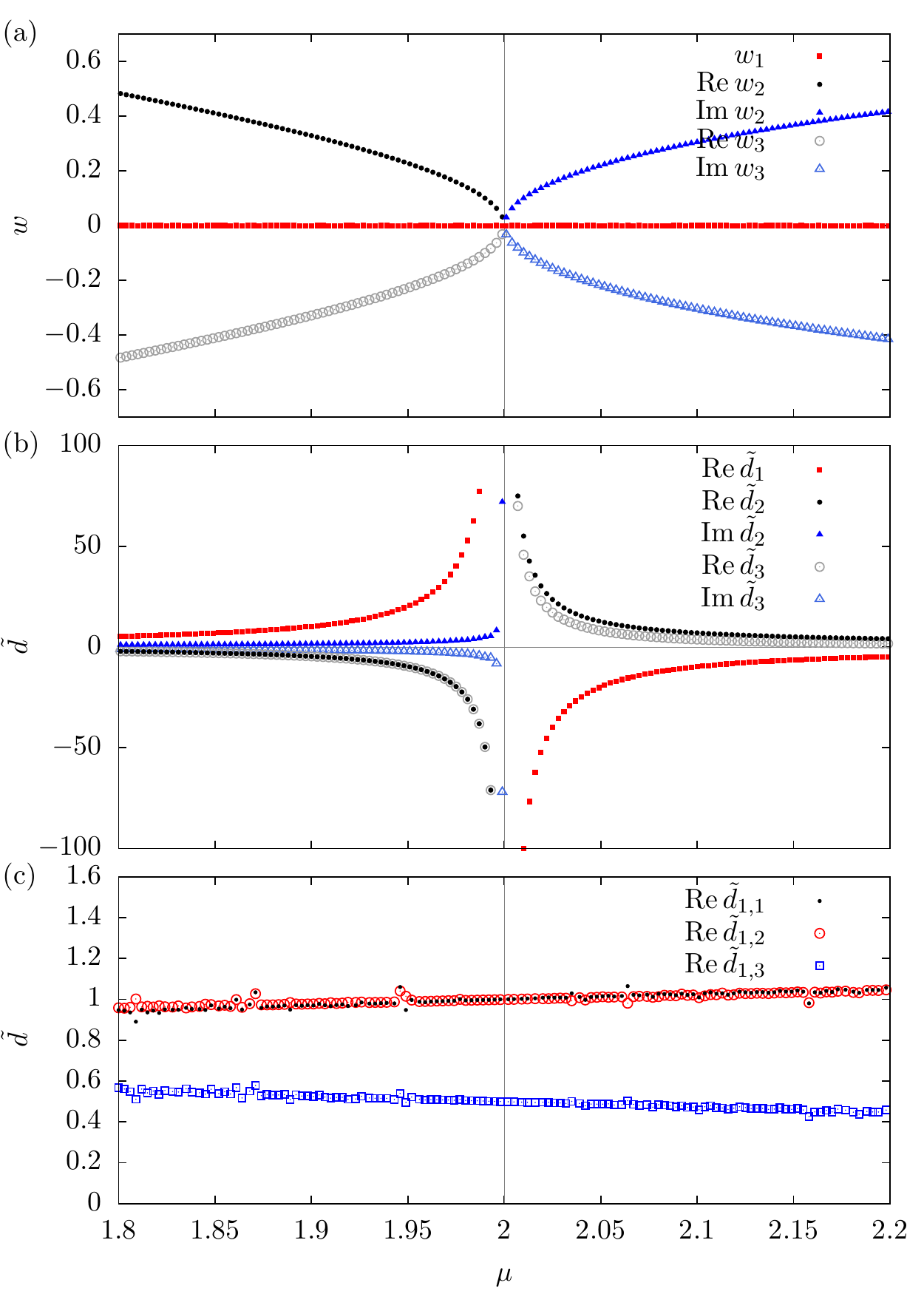}
\caption{(a) Frequencies and (b), (c) amplitudes obtained by the
  harmonic inversion analysis of the signal $C(s)$ in
  \eref{eq:C_harm_trap} as functions of the adiabatic parameter $\mu$.
  The amplitudes $\tilde d_1$ to $\tilde d_3$ in (b) obtained with
  equation \eref{eq:d_nondeg} for non-degenerate resonances diverge at
  $\mu=2$, where the three frequencies in (a) coincide.  By contrast,
  the amplitudes $\tilde d_{1,1}$ to $\tilde d_{1,3}$ in (c) obtained
  with equation \eref{eq:d_threefold} for a threefold degenerate
  resonance are smooth functions around $\mu=2$.  The nonzero value
  $\tilde d_{1,3}=1$ in the region $\mu\approx 2$ provides clear
  evidence, that the critical point is a third-order EP.}
\label{fig3}
\end{figure}

As an alternative approach we suggest to use the extended harmonic
inversion method to identify the EP unambiguously.
For the analysis of signals it is convenient to write the signal
$C(s)$ defined in \eref{eq:fouriertransformierte} in the form
\begin{equation}
  C(s) = \sum_k \sum_{\alpha=1}^{r_k} \tilde d_{k,\alpha} \,
  s^{\alpha-1} \, \exp(-\ii w_k s)
\end{equation}
with
\begin{equation}
  \tilde d_{k,\alpha} = \frac{(-\ii)^\alpha}{(\alpha-1)!} \, d_{k,\alpha} \; ,
\label{eq:tilde_d}
\end{equation}
and to determine the frequencies $w_k$ and amplitudes $\tilde d_{k,\alpha}$
with the extended harmonic inversion method.
The results for the analysis of the signal \eref{eq:C_harm_trap} are
presented in figure~\ref{fig3}.
The numerical procedure reveals that the signal possesses exactly
three frequencies with nonzero amplitudes.
The frequencies are shown in figure~\ref{fig3}(a).
Evidently, $w_1=0$ is constant and $w_2$ and $w_3$ undergo a
transition from purely real to purely imaginary frequencies at
$\mu=2$, which is directly related to the transition from an
oscillatory to a monotonic exponential dynamics in figure~\ref{fig2}.
The amplitudes $\tilde d_1$ to $\tilde d_3$ in figure~\ref{fig3}(b)
obtained with equation \eref{eq:d_nondeg} for non-degenerate
resonances diverge at $\mu=2$, where the three frequencies coincide.
By contrast, the amplitudes $\tilde d_{1,1}$ to $\tilde d_{1,3}$ in
figure~\ref{fig3}(c) obtained with equation \eref{eq:d_threefold} for
a threefold degenerate resonance are smooth functions around $\mu=2$.
The nonzero value $\tilde d_{1,3}=0.5$ in the region $\mu\approx 2$
provides clear evidence, that the critical point is a third-order EP.

In any realisation of the experiment proposed by Uzdin \etal \cite{Uzd13}
the frequency $\omega(t)$ of the time-dependent harmonic trap can
certainly be varied only within a limited finite range, and thus the
signal $C(s)$ in figure~\ref{fig2} cannot be measured at large values
$s$ of the renormalised time.
The results in figure~\ref{fig3} have been obtained by analysing the
region $s\in [0,2]$ of the signal.
Note that at small values of $s$ the signals belonging to different
parameters $\mu$ become more and more similar (see, e.g.\ the signals
with $\mu=1.8$, $\mu=2$, and $\mu=2.2$ in figure~\ref{fig2}) and thus
the extraction of the correct frequencies and amplitudes from a short
signal is a nontrivial task.
We have checked that signals with a short signal length down to
$s_{\max}\approx 1$ are sufficient to clearly observe the degeneracy
of the three frequencies at $\mu=2$ and to verify the nonzero
amplitude $\tilde d_{1,3}\approx 0.5$ indicating the third-order EP.
Furthermore, the harmonic inversion method is robust against a certain
amount of noise in the signal \cite{Bel00a,Bel00b}.
For these reasons the extended harmonic inversion method introduced in
this paper is ideally suited to reveal exceptional points in
experimentally measured signals.

\section{Conclusion}
\label{sec:conclusion}
We have extended the harmonic inversion method to allow for the
analysis of spectra and time signals with degeneracies.
In the energy or frequency domain the parameters of non-Lorentzian
line shapes related to higher order resonance poles and in the time
domain the parameters of non-exponentially decaying contributions to
time signals can be extracted.
The method has been applied to reveal exceptional points in the
photoabsorption spectrum of the hydrogen atom in crossed electric and
magnetic fields and in the dynamics of a single particle in a
time-dependent harmonic trap.
The harmonic inversion analysis is an alternative to the observation
of the permutation of states when the exceptional point is encircled
in the parameter space.
The advantage of the method is that it allows for the verification of
an EP even in cases when appropriate parameters for the encircling of
the EP are not available.
In the future the extended harmonic inversion method can be used to
observe exceptional points in a large variety of physical systems,
including e.g.\ electronic circuits, mechanical systems, atomic
spectra, Bose-Einstein condensates, microcavities, and microwave
resonators.
In particular it will be interesting to use the method for the
identification of exceptional points in systems in which they could be
applied to manipulate the system.
For example, exceptional points are discussed in molecular physics to
prepare molecules in a defined vibrational level \cite{Lef09,Gil13}.
In \cite{Die11} the EP distinguishes between the PT-symmetric and the
PT-broken phase, and in \cite{Uzd13} it is shown that an EP can be
used to shift an oscillator to an exponential decaying regime.

\appendix

\section{Derivation of the sum relation}
Here we present derivations of equations \eref{eq:sum_relation} and
\eref{eq:f_alpha} for the series $\sum_{n=0}^\infty n^{\alpha-1}x^n$.
We start with the formula
\begin{equation}
  \left(x {\textstyle\frac{\dd}{\dd x}}\right)^{\alpha} f(x)
  = \sum_{n=0}^{\alpha} \stirling{\alpha}{n} \, x^n
  \left({\textstyle\frac{\dd}{\dd x}}\right)^{n} f(x)
  \label{eq:deriv_equation}
\end{equation}
which can easily be proved by induction:
With the Stirling numbers $\stirling{\alpha}{\alpha+1}=0$
and $\stirling{\alpha}{-1}=0$ we obtain
\begin{eqnarray}
  \left(x {\textstyle\frac{\dd}{\dd x}}\right)^{\alpha+1} f(x)
  &=& \sum_{n=0}^{\alpha} \stirling{\alpha}{n}
  \left( n \, x^n \left({\textstyle\frac{\dd}{\dd x}}\right)^{n} f(x)
    + x^{n+1} \left({\textstyle\frac{\dd}{\dd x}}\right)^{n+1} f(x) \right)
  \nonumber \\
  &=& \sum_{n=0}^{\alpha+1} \left( n \, \stirling{\alpha}{n}
    + \stirling{\alpha}{n-1} \right)
  x^n \, \left({\textstyle\frac{\dd}{\dd x}}\right)^{n} f(x) \; ,
\end{eqnarray}
with an index shift $n \to n-1$ in the second term of the upper
equation.
Using the recurrence relation
$\stirling{\alpha+1}{n} = n \, \stirling{\alpha}{n} +
\stirling{\alpha}{n-1}$ \cite{Abr64}
we obtain \eref{eq:deriv_equation}.

Now this formula is applied to the geometric series.
On the one hand we have
\begin{equation}
  \left(x {\textstyle\frac{\dd}{\dd x}}\right)^{\alpha-1}
  \sum_{n=0}^{\infty} x^{n} =
  \left(x {\textstyle\frac{\dd}{\dd x}}\right)^{\alpha-1} \frac{1}{1-x}
  = \sum_{n=0}^{\alpha-1} n! \, \stirling{\alpha-1}{n} \,
  \frac{x^n}{(1-x)^{n+1}} \; .
\end{equation}
On the other hand, the derivatives of the geometric series yield
\begin{equation}
  \left(x {\textstyle\frac{\dd}{\dd x}}\right)^{\alpha-1} \sum_{n=0}^{\infty} x^{n}
  = \sum_{n=0}^{\infty} n^{\alpha-1} \, x^{n} \; ,
\end{equation}
and therefore we arrive at equation~\eref{eq:sum_relation}.
To obtain the functions $f_{\alpha}(x)$ in \eref{eq:sum_relation}
we factorise the highest powers of $x$ and $1/(1-x)$
\begin{equation}
  \sum_{n=0}^{\infty} n^{\alpha-1} \, x^{n} = \frac{x^{\alpha}}{(1-x)^{\alpha}} \,
  \sum_{n=0}^{\alpha-1} n! \, \stirling{\alpha-1}{n} \,
  \left(\frac{1}{x}\right) \left(\frac{1-x}{x}\right)^{\alpha-n-1} \; ,
\end{equation}
which yields \eref{eq:f_alpha}.

\section*{References}
\bibliographystyle{unsrt}

\end{document}